\documentclass[superscriptaddress,twocolumn,showpacs,prl,floatfix]{revtex4}

\usepackage{color}
\usepackage{tabularx}
\usepackage{epsfig}
\usepackage{amsmath}
\usepackage{amssymb}
\usepackage{graphicx}

\begin{document}

\title{An extended scaling analysis of the $S=1/2$ Ising ferromagnet
  on the simple cubic lattice}

\author{I. A.~Campbell}
\affiliation{ Laboratoire Charles Coulomb, Universit\'e Montpellier II, 34095 Montpellier, France}

\author{P. H.~Lundow}
\affiliation {Department of Theoretical Physics, Kungliga Tekniska h\"ogskolan, SE-106 91 Stockholm, Sweden}

\begin{abstract}
  It is often assumed that for treating numerical (or experimental)
  data on continuous transitions the formal analysis derived from the
  renormalization group theory can only be applied over a narrow
  temperature range, the "critical region"; outside this region
  correction terms proliferate rendering attempts to apply the
  formalism hopeless. This pessimistic conclusion follows largely from
  a choice of scaling variables and scaling expressions which is
  traditional but very inefficient for data covering wide
  temperature ranges. An alternative "extended scaling" approach can
  be made where the choice of scaling variables and scaling
  expressions is rationalized in the light of well established high
  temperature series expansion developments.
  We present the extended scaling approach in detail, and outline the
  numerical technique used to study the three-dimensional $3$d Ising model. After a
  discussion of the exact expressions for the historic $1$d Ising spin
  chain model as an illustration, an exhaustive analysis of high
  quality numerical data on the canonical simple cubic lattice $3$d
  Ising model is given. It is shown that in both models, with
  appropriate scaling variables and scaling expressions (in which
  leading correction terms are taken into account where necessary),
  critical behavior extends from $T_c$ up to infinite temperature.

\end{abstract}

\pacs{ 75.50.Lk, 05.50.+q, 64.60.Cn, 75.40.Cx}

\maketitle

\section{Introduction}
Understanding the universal critical behavior observed at and
near continuous transitions is one of the major achievements of
statistical physics; the subject has been studied in depth for many
years.  It is generally considered however that the formalism based on
the elegant renormalization group theory (RGT) can only be applied
over a narrow temperature range, the "critical region", while outside
this region correction terms proliferate so attempts to extend the
analysis become pointless. In fact this pessimistic conclusion follows
largely because the traditional choices of scaling variables and
scaling expressions are poorly adapted to the study of wide
temperature ranges.

The expressions for critical divergencies of observables $Q(T)$ near a
critical temperature $T_c$ and in the thermodynamic (infinite size)
limit are conventionally written
\begin{equation}
  Q(T) = C_{Q}t^{-q}\left(1 + F_{Q}(t)\right)
  \label{Qq}
\end{equation}
with the scaling variable $t$ defined as
\begin{equation}
  t = (T-T_c)/T_c
  \label{tdef}
\end{equation}
and where $F_{Q}(t)$ represents an infinite set of confluent and
analytic correction terms \cite{wegner:72}
\begin{equation}
  F_{Q}(t) = a_{Q}t^{\theta} + b_{Q}t + \cdots
  \label{tcorrnF}
\end{equation}
The exponents q, the confluent correction exponent $\theta$, and many
critical parameters such as amplitude ratios and finite size scaling
functions, are universal, i.e. they are identical for all members of a
universality class of systems.  When the RGT formalism is outlined in
textbooks or in authoritative reviews such as those of Privman,
Hohenberg and Aharony \cite{privman:91} or Pelissetto and Vicari
\cite{pelissetto:02} the scaling variable is defined as $t$ from the
outset.  However, because $t \to \infty$ at infinite temperature, when
$t$ is chosen as the scaling variable the correction terms in
$F_{Q}(t)$ each individually diverge as temperature is increased. It
indeed becomes extremely awkward to use the expressions in
Eq.(\ref{tcorrnF}) outside a narrow "critical" temperature region. A
"critical-to-classical crossover" has been invoked
(e.g. Refs. \cite{luijten:97,garrabos:06}) with the effective exponent
$\gamma_{\mathrm{eff}}(\beta)$ tending to the mean field values as the
high temperature Gaussian fixed point is approached. The crossover
appears as a consequence of the definition of the exponent in terms of
the thermodynamic susceptibility and the scaling variable $t$. There
is no such crossover when the extended scaling analysis described
below is used.

Although this is rarely stated explicitly, there is nothing sacred
about the scaling variable $t$; alternative scaling variables $\tau$
can be legitimately chosen and indeed have been widely used in
practice, see
e.g. Refs.~\cite{fahnle:84,gartenhaus:88,kim:96,butera:02,deng:03,caselle:97}.

Temperature dependent prefactors can also be introduced in the scaling
expressions on condition that the prefactor does not have a critical
temperature dependence at $T_c$.

An "extended scaling" approach
\cite{campbell:06,campbell:07,campbell:08,katzgraber:08,hukushima:09}
has been introduced which consists in a simple systematic rule for
selecting scaling variables and prefactors, inspired by the well
established high temperature series expansion (HTSE) method. This
approach is a rationalization which leads automatically to well
behaved high temperature limits as well as giving the correct critical
limit behavior.

Here we give a general discussion of this approach. We outline the
relationship to the RGT scaling field formalism.  As an illustration
of the application of the rules, known analytic results on the
historically important $S=1/2$ Ising ferromagnet chain in dimension
one (for which the critical temperature is of course $T_c = 0$) are
cited. Simple extended scaling expressions for the reduced
susceptibility, the second moment correlation length, and the specific
heat are exact over the entire temperature range from zero to
infinity. An exact susceptibility finite size scaling function is
exhibited.

The $S=1/2$ nearest neighbor Ising ferromagnet on the simple cubic
lattice is then discussed in detail. This model is among the principal
canonical examples of a system having a continuous phase transition at
a non-zero critical temperature.  In contrast to the two-dimensional $2$d Ising model,
in three dimensions no exact values are known for the critical temperature or the
critical exponents.  We analyze high quality large scale numerical
data which have been obtained for sizes up to $L=256$, covering wide
temperature ranges both above and below the critical temperature
\cite{haggkvist:04,haggkvist:07}. The numerical technique is outlined.
An analysis using the extended scaling approach provides compact
critical expressions with a minimum of correction terms, which are
accurate (if not formally exact) over the entire temperature range
from $T_c$ to infinity and not only within a narrow critical regime.
(The $3$d Ising, {\it XY}, and Heisenberg ferromagnets have been discussed in
Ref.~\cite{campbell:07}).

\section{Definitions of variables}

We study the $S=1/2$ nearest neighbor interaction ferromagnetic Ising
model on the $1$d chain and on the simple cubic lattices of size $L^3$
with periodic boundary conditions. The Hamiltonian with nearest
neighbor interactions of strength $J$ is
\begin{equation}
  \mathcal{H}= -J \sum_{ij} S_{i}\cdot S{j}
\end{equation}
with the sum over nearest neighbor bonds.  As usual we will use
throughout the normalized inverse temperature $\beta \equiv J/kT$.

The observables we have studied are as follows:

 (i) The variance of the equilibrium sample moment, which is equal to
 the non-connected reduced susceptibility
\begin{equation}
  \chi(\beta,L) = N\,\langle m^2 \rangle = (1/N)\,\sum_{i,j} \langle S_{i} \cdot
  S{j} \rangle
  \label{chidef}
\end{equation}
where $m$ is the magnetization per spin $m=(1/N)\sum_i S_i$, $N=L^d$.

(ii) The variance of the modulus of the equilibrium sample moment, or the
"modulus susceptibility"
\begin{equation}
  \chi_{\mathrm{mod}}(\beta,L) = N\left( \langle m^2\rangle - \langle
  |m|\rangle^2\right)
  \label{chimoddef}
\end{equation}
Below $T_c$, $\chi_{\mathrm{mod}}(\beta,L)$ tends to the connected
reduced susceptibility in the thermodynamic limit, and
\begin{equation}
  \langle |m| \rangle(\beta,L) =
  \sqrt{\chi(\beta,L)-\chi_{\mathrm{mod}}(\beta,L)}/\sqrt{N}
  \label{mabsdef}
\end{equation}
tends to the thermodynamic limit magnetization $\langle m
\rangle(\beta,L)$ at large $L$.

(iii) The specific heat which is equal to the variance of the energy
per spin $U(\beta,L)$
\begin{equation}
  C_{v}(\beta,L) = N\,\left(\langle U^2\rangle - \langle U\rangle^2\right)
  \label{Cvdef}
\end{equation}
where U = $(1/N)\,\sum_{ij} S_i\cdot S_j$ with the sum over nearest
neighbor bonds.  We can note that $\chi(\beta)$ and $C_v(\beta)$ have
consistent statistical definitions in terms of thermal
fluctuations. The experimentally observed susceptibility contains an
extraneous factor $\beta$.

The thermodynamic limit second moment correlation length is defined
\cite{fisher:67,butera:02} by
\begin{equation}
  \xi^2(\beta,\infty) = \mu_2(\beta,\infty)/2d\chi(\beta,\infty)
  \label{ximu2def}
\end{equation}
where the second moment of the correlation function is
\begin{equation}
  \mu_2(\beta,\infty) = (1/N)\sum_{i,j}\langle r_{i,j}^2 S_{i} \cdot S_{j} \rangle
  \label{mu2def}
\end{equation}
with $r_{i,j}$ the distance between spins $i$ and $j$, summing to infinity.

When the "thermodynamic limit" condition $L \gg \xi(\beta,\infty)$
holds all properties become independent of $L$ and so are identical to
the thermodynamic limit properties.

For general $L$, the Privman-Fisher finite size scaling {\it ansatz}
for an observable $Q$ can be written \cite{brezin:82,calabrese:03}
\begin{eqnarray}
  Q(\beta,L)/Q(\beta,\infty) = & \\ \nonumber
  F_{Q}\left(L/\xi(\beta,\infty)\right) &
  \left(1 + L^{-\omega} G_{Q}\left(L/\xi(\beta,\infty)\right)\right)
  \label{PFFSS}
\end{eqnarray}
The functions $F_{Q}\left(x\right)$, $G_{Q}\left(x\right)$ are
universal. $F_{\chi}\left(x\right)$ must tend to $1$ when $x \gg 1$,
and must be proportional to $x^{2-\eta}$ when $x \ll 1$. We are aware
of no generally accepted explicit expressions for
$F_{Q}\left(x\right)$ valid over the entire range of $x$.

\section{Extended scaling}

In the extended scaling approach \cite{campbell:06,campbell:07} a
systematic choice of scaling variables and scaling expression
prefactors is made in the light of the HTSE. Basically, an ideal HTSE
corresponds to the power series
\begin{eqnarray}
  (1-y)^{-q} \equiv 1 + qy + (q(q+1)/2)y^2 +\cdots
  \label{darboux}
\end{eqnarray}
When a real physical HTSE has the form
\begin{equation}
  Q\left(x\right) = C_Q\,\left(1+a_1x+a_2x^2+\cdots\right)
  \label{HTSE}
\end{equation}
with a general structure similar to but not strictly equivalent to
that of Eq.(\ref{darboux}) and a prefactor $C_Q$ which can be
temperature dependent, the asymptotic limit is eventually dominated by
the closest singularity to the origin (Darboux's first theorem
\cite{darboux:78}) leading to the critical limit $Q(x)= C_{Q}(1 -
x)^{-q}$. The appropriate critical scaling variable is $1-x$, and
deviations of the series in Eq.(\ref{HTSE}) from the pure
Eq.(\ref{darboux}) form correspond to confluent and analytic critical
correction terms.

The extended scaling prescription consists in identifying scaling
variables and prefactors such that each series is transposed to a form
having the same structure as Eq.(\ref{HTSE}), with the prefactor
defined so that the first term of the series is equal to $1$.

The HTSE spin $S=1/2$ expressions for the reduced susceptibility and
the second moment of the correlation can be written generically in the
form \cite{fisher:67,butera:02,butera:lib}
\begin{equation}
  \chi(\beta) = 1 + a_{1}x + a_{2}x^2 + a_{3}x^3 + \cdots
  \label{chiHTSE}
\end{equation}
and
\begin{equation}
  \mu_{2}(\beta) =  b_{1}x + b_{2}x^2 + b_{3}x^3 + \cdots
  \label{mu2HTSE}
\end{equation}
where $x$ is a normalized variable which tends to $1$ as $\beta \to
\beta_c$ and to zero when $\beta \to 0$.

For ferromagnets, (e.g.~\cite{fisher:67,butera:02,butera:lib})
possible natural choices for $x$ are $x=\beta/\beta_c$ or
$x=\tanh\beta/\tanh\beta_c$. Scaling variables for $\chi(\beta)$
are $\tau = 1-x = 1- \beta/\beta_c$ or $\tau= 1- x = 1-
\tanh\beta/\tanh\beta_c$. The former is standard when $T_c$ is
non-zero; when $T_c = 0$, it is convenient to use $x = \tanh\beta$
(as $\beta_c = \infty$, $\tanh\beta_c =1$).

For $\mu_2(\beta)$ the Eq.(\ref{HTSE}) form with the same $x$ can be
retrieved by extracting a temperature dependent prefactor $b_{1}x$ so
as to write
\begin{equation}
  \mu_2(\beta) = b_{1}x(1 + (b_{2}/b_{1})x + (b_{3}/b_{1})x^2 +\cdots)
\end{equation}

The critical expressions for the reduced susceptibility and the second
moment correlation length can then be written
\begin{equation}
  \chi(\beta,\infty) = C_{\chi}\tau^{-\gamma}\left( 1 + F_{\chi}(\tau) \right)
  \label{chiextdef}
\end{equation}
(c.f. Eq.(\ref{Qq})) and from the relation Eq.(\ref{ximu2def}) between
$\mu_2$ and $\xi$,
\begin{equation}
  \xi(\beta,\infty) = C_{\xi}x^{1/2}\tau^{-\nu}\left( 1 +
  F_{\xi}(\tau)\right)
  \label{xiextdef}
\end{equation}
with the temperature scaling variable $\tau = 1-x$ and the standard definitions for the critical amplitudes $C_{\chi}$ and $C_{\xi}$.
The $\chi(\beta,\infty)$ expression has been widely used; the
$\xi(\beta,\infty)$ expression is specific to the extended scaling
approach \cite{campbell:06,campbell:07}.  The $F$ functions contain
all the confluent and analytic correction to scaling terms
\cite{wegner:72,wegner:76}
\begin{equation}
  F_{Q}(\tau) = a_{Q}\tau^{\theta} + b_{Q}\tau + \cdots
  \label{wegnertau}
\end{equation}
It is important that $\tau$ tends to $1$ at infinite temperature
(whereas $t$ tends to infinity); the $F_{Q}(\tau)$ thus remain well
behaved over the entire temperature range.  There are exact closure
conditions for the infinite temperature limit $\tau \to 1$ :
$C_{\chi}(1 + F_{\chi}(1)) = 1$ and $C_{\xi}/\beta_c^{1/2}(1 +
F_{\xi}(1)) = 1$ (or $C_{\xi}/(\tanh\beta_c)^{1/2}(1 + F_{\xi}(1)) =
1$).

One can define temperature dependent effective exponents (introduced
by \cite{kouvel:64}):
\begin{equation}
  \gamma_{\mathrm{eff}}(\tau)= \partial{\log\chi(\beta)}/\partial{\log\tau}
  \label{gammaeff}
\end{equation}
see Refs.~\cite{orkoulas:00,butera:02}. For the correlation length,
\begin{equation}
  \nu_{\mathrm{eff}}(\tau) =
  \partial{\log(\xi(\beta)/\beta^{1/2})}/\partial{\log\tau}
  \label{nueff}
\end{equation}
is the extended scaling definition for $\nu_{\mathrm{eff}}$.

For a spin $S=1/2$ Ising ferromagnet on a lattice where each spin has
$z$ neighbors, the high temperature limit of the effective exponents
defined by Eqns.~\ref{gammaeff} and \ref{nueff} are
$\gamma_{\mathrm{eff}}(1) = z\beta_c$ and $\nu_{\mathrm{eff}}(1)=
\gamma_{\mathrm{eff}}(1)/2$. A comparison between these values and the
critical exponents $\gamma$ and $\nu$ gives a good indication of the
overall influence of the correction terms. If the leading confluent
correction term in Eq.(\ref{wegnertau}) dominates then
$\gamma_{\mathrm{eff}}(1)-\gamma \approx a_{\chi}\theta$, and
$\nu_{\mathrm{eff}}(1)-\nu \approx a_{\xi}\theta$. An analysis along
these lines of $\gamma_{\mathrm{eff}}$ for Ising systems with large
$z$ was sketched out in Ref.\cite{orkoulas:00}. The case of general
$S$ is discussed in Appendix A. For all near neighbor Ising ferromagnets on sc or
bcc lattices covering the entire range of spin values $S=1/2$ to
$S=\infty$ (which are all in the same $3$d universality class), see
Ref.~\cite{butera:02}, $\gamma_{\mathrm{eff}}(1)$ and
$\nu_{\mathrm{eff}}(1)$ differ from the critical $\gamma$ and $\nu$ by
a few percent at most. For both observables, the total sum of the
correction terms is weak over the entire temperature range.

It should be noted that traditional and widely used finite size
scaling expressions
\begin{equation}
  Q(\beta,L)/L^{q/\nu} = F_{Q}\left( L^{1/\nu}(T-T_c)/T_c \right)
  \label{tradFSS}
\end{equation}
assume implicitly scaling with the scaling variable $t$.  As a general
rule these expressions should not be used except in the limit of
temperatures very close to $T_c$; they rapidly becomes misleading and
can suggest incorrect values of the exponent $\nu$ if global fits are
made to data covering a wider range of temperatures. The extended
scaling FSS expressions \cite{campbell:06}
\begin{equation}
  Q(\beta,L)/(LT^{1/2})^{q/\nu} =
  F_{Q}\left((LT^{1/2})^{1/\nu}(T-T_c)/T \right)
  \label{extFSS}
\end{equation}
are valid at all temperatures above $T_c$ to within the weak
corrections to scaling.

For spin $S=1/2$ Ising spins on a bipartite lattice (such as the $1$d
and $3$d sc lattices we will discuss below) there are only even terms
in the HTSE for the specific heat \cite{butera:02,butera:lib}
\begin{equation}
  C_v(\beta) = 1 + d_{1}x^2 + d_{2}x^4 + d_{3}x^6 + \cdots
\end{equation}
A natural scaling expression for the specific heat is
\begin{equation}
  C_v(\beta) = C_{0} + C_{c}\left( 1-x^2 \right)^{-\alpha}\left( 1 +
  F_{c}(1-x^2) \right)
  \label{Cvextdef}
\end{equation}
The constant term $C_0$ is present in standard analyses and plays an
important r\^ole in $3$d ferromagnets because the exponent $\alpha$ is
small. The extended scaling expression Eq.(\ref{Cvextdef}) is not
orthodox as it uses a scaling variable, $\tau_2 = 1-x^2
=\tau(2-\tau)$, which is not the same as the $\tau = 1 - x$ used as
scaling variable for $\chi(\beta)$ and $\xi(\beta)$.

\section{Analytic results in one dimension}

The original Ising ferromagnet \cite{ising:25} consists of a system of
$S=1/2$ spins with nearest neighbor ferromagnetic interactions on a
one dimensional chain. Because analytic results exist for many of the
statistical properties of this system, it is often used as a
"textbook" model in introductions to critical behavior. We will use it
to illustrate the extended scaling approach
(see Ref. \cite{katzgraber:08}).

The model orders only at $T = 0$ (Ref. \cite{ising:25}); when $T_c=0$ the
critical exponents depend on the choice of the scaling
variable. Baxter \cite{baxter:82} states : "[in one dimension] it is more
sensible to replace $t=(T-T_c)/T_c$ by $t = \exp(-2\beta)$"; with this
scaling variable the exponents are $\gamma = 1, \nu = 1, \alpha = -1$
[$\alpha = -\nu d$ when $T_c = 0$ Ref. \cite{baker:75}].

Expressions for $\xi(\beta)$ and $\chi(\beta)$ in the infinite-size
limit are readily calculated following standard HTSE rules (see
e.g. Ref.~\cite{fisher:67}).  The reduced susceptibility HTSE can be
written as
\begin{equation}
  \chi(\beta) = 1 + 2(\tanh\beta + \tanh^2\beta + \tanh^3\beta +
  \ldots)
\end{equation}
and the HTSE for the second-moment of the correlation is
\begin{equation}
  \mu_2(\beta) = 2(\tanh\beta + 2^2\tanh^2\beta + 3^2\tanh^3\beta + \ldots)
\end{equation}
The second-moment correlation length is then given by
Eq.~(\ref{ximu2def}) with $z = 2$. Using the power series sums
\begin{equation}
  \sum_{n=1}^{\infty} y^n = \frac{y}{(1-y)}
\end{equation}
and
\begin{equation}
  \sum_{n=1}^{\infty} n^2y^n =  \frac{y(y+1)}{(1-y)^3}
\end{equation}
the exact expressions for reduced susceptibility and correlation
length are thus
\begin{equation}
  \chi(\beta)  = \exp(2\beta)
  \label{chi1d}
\end{equation}
and
\begin{eqnarray}
  \xi(\beta) &=& \frac{1}{2}\left(\exp(4\beta)-1\right)^{1/2} \nonumber \\	 
    \label{xi1d}
\end{eqnarray}
(It can be noted that the "true" correlation length is $\xi_{\mathrm{true}} =
-1/\log(\tanh\beta)$. The two correlation lengths are essentially
identical for $\beta > 2.5$ but are quite different at higher
temperatures.)

The internal energy per spin is just
\begin{equation}
  U(\beta)= \tanh\beta
\end{equation}
so the specific heat
\begin{equation}
  C_v(\beta)= \cosh^{-2}\beta
\end{equation}

Though not immediately recognizable these can all be re-written in
precisely the form of the extended scaling
Eqns.~\ref{chiextdef},\ref{xiextdef},\ref{Cvextdef}, with the choice
$x = \tanh\beta$ so $\tau=(1-\tanh\beta)$;
\begin{equation}
  \chi(\beta) = 2(1-\tanh\beta)^{-1}\left( 1 - (1/2)(1-\tanh\beta)\right)
  \label{chiext1d}
\end{equation}
\begin{eqnarray}
  \xi(\beta) = \frac{\tanh^{1/2}\beta}{1 - \tanh\beta}
  \label{xiext1d}
\end{eqnarray}
and
\begin{equation}
  C_v(\beta,\infty)=  (1 - \tanh^2\beta)
\label{Cvext1d}
\end{equation}
and so with the same critical exponents $\gamma = 1$, $\nu = 1$,
$\alpha = -1$ together with critical amplitudes $C_{\chi} = 2$,
$C_{\xi} = 1$, $C_v = 1$, $C_0 = 0$. There are no analytic corrections
to $\xi(\beta)$ or to $C_v(\beta)$ and there is only a single simple
analytic correction to $\chi(\beta)$. There are no confluent
corrections.  Note again that these expressions are valid for the {\em
  entire} temperature range from $T=0$ to $T=\infty$.

The finite size scaling function can also be considered.  With
periodic boundary conditions the finite size reduced susceptibility
for a $1$d sample of size $L$ is
\begin{equation}
  \chi(\beta,L) = \exp(2\beta)\frac{1 - \tanh^L\beta}{1 + \tanh^L\beta}
\end{equation}
The finite size scaling function is
\begin{eqnarray}
  \chi(\beta,L)/\chi(\beta,\infty)~~~~~~~~~~~~ \\ \nonumber =
  \tanh(L/2\xi(\beta,\infty))(1+ L^{-2}G_{\chi}\left(
  L/2\xi(\beta,\infty)\right)
  \label{FSS1d}
\end{eqnarray}
The simple principle expression
\begin{equation}
  F_{\chi}(L/\xi(\beta,\infty)) = \tanh(L/2\xi(\beta,\infty))
\end{equation}
is exact.

The higher order term $G_{\chi}(x)$ in Eq.(\ref{FSS1d}) is
numerically tiny even for small $L$. We have not found an analytic
expression but it can be fitted rather accurately by
\begin{equation}
  G_{\chi}(x) = 0.168x^2\left( 1+\tanh(-0.565x^{1.18})\right)
\end{equation}

\section{High dimension limit}

For the Ising ferromagnet in the high dimension hypercubic lattice
limit $d \to \infty$, with $\tau(\beta) = 1-\beta/\beta_c$ the reduced
susceptibility and the correlation length are
\begin{equation}
  \chi(\beta,\infty) \equiv \tau^{-1}
  \label{chidinf}
\end{equation}
and
\begin{equation}
  \xi(\beta,\infty) \equiv (\beta/\beta_c)^{1/2}\tau^{-1/2}
  \label{xidinf}
\end{equation}
exactly over the entire temperature range above $T_c$; the exponents
$\gamma = 1, \nu = 1/2$ are of course the mean field exponents. These
expressions again follow the extended scaling form given above
Eqns.~\ref{chiextdef},\ref{xiextdef} including the square root
prefactor in $\xi(\beta,\infty)$, with no correction terms.

In this high dimension limit the specific heat above $T_c$ is zero ($\alpha=0$).

In dimensions above the upper critical dimension but not in the
extreme high dimension limit the extended scaling approach has been
used successfully to identify the main correction terms in the reduced
susceptibility \cite{berche:08}.

Thus analytic expressions for models both in the low ($1$d) and high
($d \to \infty$) dimension limits follow the extended scaling
forms. This reinforces the argument that these forms can be considered
to be generic and should be used at leading order also for
intermediate dimensions, where confluent correction terms and small
analytic correction terms must be allowed for.

In practice
(e.g. Refs.~\cite{deng:03,kim:96,gartenhaus:88,caselle:97,orkoulas:00})
analyses of $\chi(\beta)$ have long been carried out using $\tau$ as
the scaling variable rather than $t$.  There are analogous advantages
in scaling $\xi(\beta)$ with Eq.(\ref{xiextdef}), which contains the
generic $(\beta/\beta_c)^{1/2}$ (or
$(\tanh\beta/\tanh\beta_c)^{1/2}$) prefactor. We suggest that this
form of scaling expression for $\xi(\beta)$ could profitably become
equally standard.

\section{RGT formalism and extended scaling}

In the standard RGT finite size scaling formalism
\cite{privman:91,pelissetto:02} the free energy is written
\begin{equation}
  \mathcal{F}(\beta,h,L) =
  \mathcal{F}_{\mathrm{sing}}(\beta,h,L) +
  \mathcal{F}_{\mathrm{reg}}(\beta,h,L)
  \label{freeenergy}
\end{equation}
where the singular part encodes the critical behavior and the regular
part is practically $L$ independent.  Then
\begin{eqnarray}
  \mathcal{F}_{\mathrm{sing}} = L^{-d} F\left(
  u_{h}L^{(d+2-\eta)/2},u_{t}L^{1/\nu}\right) + \\ \nonumber
  v_{\omega}L^{-(d+\omega)}F_{\omega}\left(
  u_{h}L^{(d+2-\eta)/2},u_tL^{1/\nu}\right) +\cdots
  \label{Fsing}
\end{eqnarray}
with the scaling fields $u_h$ and $u_t$ having temperature dependencies
\begin{equation}
  u_h = ha_{h}(1 + a_{1}t + a_{2}t^2...)
  \label{uhdef}
\end{equation}
and
\begin{equation}
  u_t = t(1 + c_{1}t + c_{2}t^2 +...)
  \label{utdef}
\end{equation}
where $h$ is the magnetic field. The two series in $t$ are analytic.

Ignoring for the moment the confluent correction series
$F_{\omega}$, for phenomenological couplings $R$
\begin{eqnarray}
  R(\beta,L) = F_{R}\left( u_{t}L^{1/\nu}\right)~~~~~~ \\ \nonumber
  = G_{R}\left( L^{1/\nu}t(1 + c_{1}t + c_{2}t^2 +\cdots)\right)
  \label{Rscal}
\end{eqnarray}
and for $\chi$
\begin{eqnarray}
  \chi(\beta,L) = AL^{2-\eta}\left( 1 + b_{1}t + b_{2}t^2 +\cdots \right) \\
  \nonumber
  = G_{\chi}\left( L^{1/\nu}t(1 + a_{1}t + a_{2}t^2 +\cdots)\right)
  \label{chiscal}
\end{eqnarray}
Analyses using this formalism are carried out by introducing a
series of analytic terms in powers of $t$, adjusting for each
particular case the constants $a_{n}, b_{n}$ and $c_{n}$ and
truncating at some power of $t$.

Now consider the extended scaling scheme. As a first step
$t=(T-T_c)/T_c$ is replaced in the formalism by $\tau=(T-T_c)/T$ just
as for instance in \cite{hasenbusch:08}. The variable $t$ is replaced
by $\tau(1-\tau)$ everywhere. This leaves the generic form of the
equations unchanged but modifies the individual factors in the series
for the temperature dependencies of the scaling fields.

In the extended scaling approach a second step must then be made due
to the $(\beta/\beta_c)^{1/2}$ prefactor in $\xi(\beta,\infty)$.  The
extended scaling FSS expressions \cite{campbell:06}
\begin{equation}
  R(\beta,L) = F_{R}\left(\tau L^{1/\nu}\beta^{-1/2\nu}\right)
  \label{Rextscal}
\end{equation}
and
\begin{equation}
  \chi(\beta,L) = (L\beta^{1/2})^{2-\eta} F_{\chi}\left(\tau
    L^{1/\nu}\beta^{-1/2\nu}\right)
  \label{chiextscal}
\end{equation}
can be translated into the RGT FSS formalism in terms of explicit
built-in leading expressions for the temperature variation of the
scaling fields.  The extended scaling expressions without correction
terms are strictly equivalent to leading expressions for the scaling
fields $u_{t}$ and $u_{h}$ containing specific infinite analytic
series of terms in $\tau^n$ :
\begin{eqnarray}
  u_{t} \sim \tau(1 - \tau)^{-1/2\nu}   \\   \nonumber
  = \tau\left( 1 + \frac{1}{2\nu}\tau + \frac{(1+2\nu)}{8\nu^2}\tau^2
  +\cdots \right)
  \label{utextscal}
\end{eqnarray}
and
\begin{eqnarray}
  u_{h} \sim h(1-\tau)^{-(2-\eta)/2}      \\   \nonumber
  = h\left( 1 + \frac{(2-\eta)}{2}\tau
  -\frac{(2-\eta)(4-\eta)}{8}\tau^2 +\cdots \right)
  \label{uhextscal}
\end{eqnarray}
In the extended scaling approach these leading expressions are
common to all ferromagnets.  The confluent correction contributions
will of course still exist with the confluent correction terms
expressed using $\tau$. Finally, fine tuning through minor
modifications of the analytic scaling field temperature dependence
series will usually be necessary to obtain higher level approximations
to the overall temperature variation of the observables.

Not only at temperatures well above $T_c$ but already at criticality
the extended scaling scheme can aid the data analysis. For instance,
quite generally the critical size dependence of the ratio of the
derivative of the susceptibility to the susceptibility is of the form
\cite{hasenbusch:08,katzgraber:06}
\begin{equation}
  (\partial{\chi(\beta,L)}/\partial{\beta})/\chi(\beta,L) =
  K_{1}L^{1/\nu}\left( 1 + c_{\omega}L^{-\omega}+ \cdots \right) +
  K_{2}
  \label{dchidbeta}
\end{equation}
An explicit leading order value of the $L$-independent term can be
derived from the leading order extended scaling FSS
Eq.(\ref{chiextscal}):
\begin{equation}
  K_{2} = -(2-\eta)/2\beta_c
  \label{dchidbetaK2}
\end{equation}
in a ferromagnet. This value will be slightly modified by a correction
to scaling term.

The extended scaling scheme can thus be translated unambiguously into
the standard RGT FSS formalism.  It can be considered as providing an
{\it a priori} rationalization giving explicit leading analytical
temperature dependencies of the scaling fields. At this level the
extended scaling scheme provides compact baseline expressions which
cover the entire temperature region from $T_c$ to infinity, accurate
to within confluent correction terms and residual model dependent
analytic correction terms.

\section{Numerical methods applied to the simple cubic model}

The equilibrium distributions of the parameters energy $p(U)$ for
finite size samples from $L=4$ up to $L=256$ ($16,777,216$ spins) were
estimated using a density of states function method. When studying a
statistical mechanical model complete information can in principle be
obtained through the density of states function. From complete
knowledge of the density of states one can immediately work with the
microcanonical (fixed energy) ensemble and of course also compute the
partition function and through it have access to the canonical (fixed
temperature) ensemble as well. The main problem here is that computing
the exact density of states for systems of even modest size is a very
hard numerical task. However, several sampling schemes have been given
for obtaining approximate density of states, of which the best known
are the Wang-Landau \cite{wang:01} and Wang-Swendsen \cite{wang:02}
methods. In \cite{haggkvist:04} the various methods are described
along with an improved histogram scheme. For work in the
microcanonical ensemble the sampling methods give all the information
needed. Using them one can find the density of states in an energy
interval around the critical region and that is all that is required
for most investigations of the critical properties of the model.

For the present analysis a density of states function technique based
upon the same method as in \cite{haggkvist:07} was used though with
considerable numerical improvements for all $L$ studied here (adequate
improvements to the $L=512$ data set would unfortunately have been too
time-consuming). The microcanonical (energy dependent) data were
collected as described in \cite{haggkvist:04}. We use standard
Metropolis single spin-flip updates, sweeping through the lattice
system in a type-writer order. Measurements take place when the
expected number of spin-flips is at least the number of sites. For
high temperatures this usually means two sweeps between measurements
and three or four sweeps for the lower temperatures we used. Note that
in the immediate vicinity of $\beta_c$ the spin-flip probability is
very close to $50\%$ for the $3$d simple cubic lattice.

We report here data on the $3$d simple cubic $S=1/2$ Ising model with
periodic boundary conditions.  For $L=256$, the largest lattice
studied here, we have now amassed between 500 and 3500 measurements on
an interval of some 450000 energy levels, where most samplings are
near the critical energy $U_c$.  For $L=128$ we have between 5000 and
50000 measurements on some 150000 energy levels.  For $L\le 64$ the
number of samplings are of course vastly bigger.

Our measurements at each individual energy level include local energy
statistics and magnetization moments.  The microcanonical data were
then converted into canonical (temperature dependent) data according
to the technique in \cite{lundow:09}. This gave us energy
distributions from which we obtain energy cumulants (e.g. the specific
heat) and together with the fixed-energy magnetization moments we
obtain magnetization cumulants (e.g. the susceptibility).

Typically around 200 different temperatures were chosen to compute
these quantities, with a somewhat higher concentration near $\beta_c$
particularly for the larger $L$ so that one may use standard
interpolation techniques on the data to obtain intermediate
temperatures.

Below $T_c$ the variance of the distribution of $m$ in zero field,
Eq.(\ref{chidef}), represents the non-connected susceptibility; the
physical susceptibility in the thermodynamic limit is the connected
susceptibility
\begin{equation}
  \chi_{\mathrm{conn}}(\beta, L) = N \left(
  \langle m^2 \rangle - (\langle m \rangle)^2\right)
  \label{connchidef}
\end{equation}
For finite $L$ the distribution of $m$ below $T_c$ is bimodal but
always symmetrical so in zero applied field $\langle m \rangle = 0$,
which would suggest that supplementary measurements are needed using
small applied fields in order to estimate
$\chi_{\mathrm{conn}}$. However under the condition $L >>
\xi_{conn}(\beta,\infty)$, where $\xi_{conn}(\beta,\infty)$ is the second moment correlation length below $T_c$, the two peaks in the distribution of $m$ become very well separated and the variance of the distribution of the
absolute value $|m|$ can be taken as essentially equal to the
connected susceptibility,
\begin{equation}
  \chi_{\mathrm{conn}}(\beta, L) = \chi_{\mathrm{mod}}(\beta, L)
\end{equation}
The explicit expression for $\xi_{conn}$ is complicated, see \cite{arisue:95}, but the onset of thermodynamic limit conditions can judged by inspection of the finite size $\chi_{\mathrm{mod}}(\beta, L)$ data.
To estimate the ordering temperature $T_c$ we have used the size
dependence of $U_4(\beta,L)$ the kurtosis of the distribution of
$p(m)$, frequently expressed in terms of the Binder parameter
$g(\beta,L)$.

We have introduced \cite{lundow:10} an alternative parameter with the
same formal properties as $g(\beta,L)$ which involves
$\chi_{\mathrm{mod}}$. The normalized parameter $W(\beta,L)$ is
defined by
\begin{equation}
  W(\beta,L) = 1-\pi(\chi_{\mathrm{mod}}(\beta,L)/\chi(\beta,L))/(\pi-2)
\end{equation}
or
\begin{equation}
  W(\beta,L) =
    \left(\pi\left(\langle |m| \rangle^2/\langle m^2\rangle\right)-2\right)
  /\left(\pi-2\right)
  \label{Wdef}
\end{equation}
The normalization has been chosen such that, as for the Binder
parameter, $W = 0$ in the high temperature Gaussian limit and $W = 1$
in the low temperature ferromagnetic limit. As $W(\beta,L)$ is also a
parameter characteristic of the shape of the distribution $p(m)$, it
can be considered to be another "phenomenological coupling". It turns
out that at least for the $3$d Ising model the corrections to scaling
for $W(\beta,L)$ are much weaker than those for $g(\beta,L)$, allowing
accurate estimates of $T_c$ and $\nu$ from scaling at criticality. The
values estimated for the critical parameters $\beta_c$ and $\nu$ are
in good agreement with the most accurate values from RGT, HTSE, and
Monte Carlo methods \cite{lundow:10}.

\section{$3$d Ising ferromagnet susceptibility and correlation length}

The Ising ferromagnet in dimension three is a canonical example of a
system having a continuous phase transition at a non-zero critical
temperature. In $3$d there are no observables which diverge
logarithmically in contrast to the $2$d and $4$d models.  Though there
are no exact results for this universality class, rather precise
estimates of the critical exponents (and the critical temperatures)
have been obtained and improved over the years thanks to extensive
analytical, HTSE, and FSS Monte Carlo studies. The essential aim has
been to determine as accurately as possible the universal critical
parameters.

Consider first the finite size scaling results at and very close to
the critical temperature. The numerical work \cite{lundow:10} provided
an estimate $\beta_c = 0.2216541(2)$ from intersections between curves
for phenomenological couplings at different sizes $L$, using data on
the Binder cumulant $g(L)$ and on the phenomenological coupling $W(L)$
\cite{lundow:10}. This value is consistent with the Monte Carlo
estimates $\beta_c = 0.22165452(8)$ \cite{deng:03} and $\beta_c =
0.22165463(8)$ \cite{hasenbusch:10a}, the HTSE estimate $\beta_c =
0.221655(2)$ \cite{butera:02}, and $\beta_c= 0.2216546(3)$
\cite{haggkvist:07}.

At criticality, the standard FSS expression \cite{salas:00} for
$\chi(\beta_c,L)$ is
\begin{equation}
  \chi(\beta_c,L) = C^{'}_{\chi} L^{2-\eta}\left( 1 +
  a^{'}_{1}L^{-\omega} + b^{'}_{1}L^{-\omega_2} +
  a^{'}_{2}L^{-2\omega} + \cdots \right)
\label{chicrit}
\end{equation}
For the $3$d Ising ferromagnet, $\theta = 0.504(8)$ or
$\omega=\theta/\nu = 0.800(13)$ \cite{guida:98} so $2\omega =
1.60(3)$. The subleading irrelevant exponent is $\omega_2 = 1.67(11)$
\cite{newman:84} so the $\omega_2$ and $2\omega$ terms can be treated
together as a single effective term $b^{'}_{2}L^{-1.65}$. In what
follows we will assume for convenience $\theta = 0.50$.

\begin{figure}
  \includegraphics[width=3.5in]{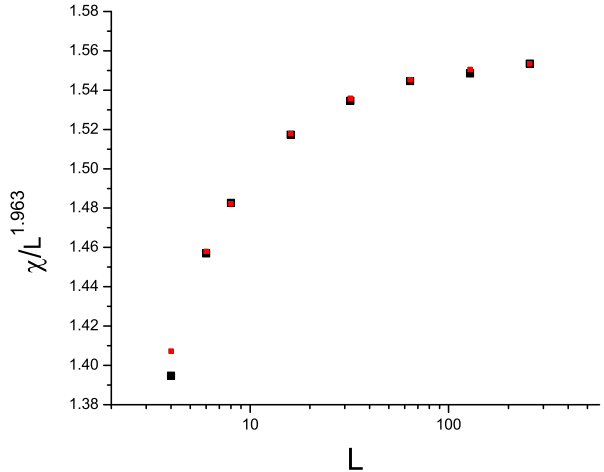}
  \caption{(Color online) Finite size corrections at the critical
    temperature. $\chi(\beta_c,L)/L^{2-\eta}$ against $L$ at $\beta_c$
    adopting $\eta=0.0368$. The large black points are measured; the small red points are the fit, Eq.(\ref{chicrit2})} \protect\label{fig:1}
\end{figure}

Fig. 1 shows $\chi(\beta_c,L)/L^{2-\eta}$ against $L$ adopting
$\eta=0.0368$ \cite{deng:03}; the finite size scaling corrections in
the present data can be fitted by
\begin{equation}
  \chi(\beta_c,L) = 1.557 L^{1.9632}\left( 1 - 0.218L^{-0.82} -
  0.256L^{-1.65}\right)
  \label{chicrit2}
\end{equation}
The analysis is consistent with that of \cite{deng:03},
\begin{equation}
  \chi(\beta_c,L) = L^{2-\eta}\left( 1.559(16)-0.37(5)L^{-0.8} \right)
\end{equation}
Because of the introduction of a next to leading term, the fit extends
to lower $L$.

\begin{figure}
  \includegraphics[width=3.5in]{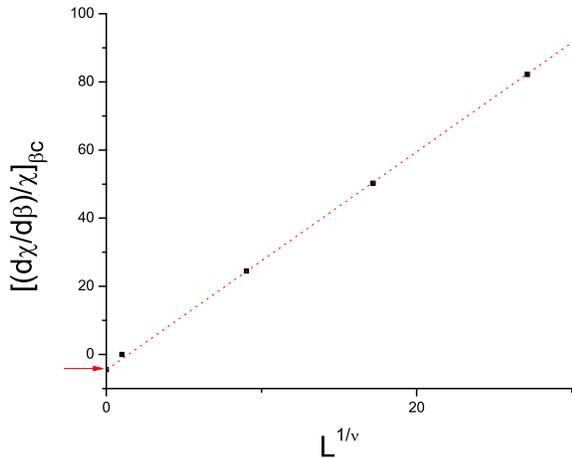}
  \caption{(Color online) The normalized derivative of the
    susceptibility $
    [(\partial\chi(\beta,L)/\partial\beta))/\chi(\beta,L)]_{\beta_c}$
    against $L^{1/\nu}$. The extended scaling value for the intercept is
    $-(2-\eta)/2\beta_c$ to leading order (red
    arrow)}\protect\label{fig:2}
\end{figure}

Fig. 2 shows partial data for the ratio
\begin{equation}
  x(L) = [(\partial\chi(\beta,L)/\partial\beta)/\chi(\beta,L)]_{\beta_c}
\end{equation}
against $L$.  On this scale the data can be well represented
by $x(L) = -(2-\eta)/2\beta_c + K_{1}L^{-1/\nu}$ with $\beta_c =
0.2216549$, $\eta = 0.0368(2)$, $\nu = 0.6302(1)$, and $K_{1}$ a
constant, see the extended scaling expression
Eq.(\ref{dchidbeta}). This form of plot provides an independent
estimate for $\nu$ consistent with the values given in
\cite{butera:02,deng:03,lundow:10}. To obtain an accurate value for
$\nu$ it is important to include the non-zero intercept.

Combining $\nu$ and $\eta$ estimates from FSS at criticality, the
present data are almost consistent with the MC and HTSE estimates
$\gamma = (2-\eta)\nu = 1.2372(4)$ \cite{deng:03} and $\gamma =
1.2371(1)$ \cite{butera:02}. Both of these are from meta-analyses on
many systems in the same universality class, the latter relying
principally on bcc data. A recent very precise study of the 3d Ising
universality class \cite{hasenbusch:10a} gave $\nu = 0.63002(10)$ and $\eta = 0.03627(10)$ so $\gamma =1.2372(3)$ together with $\omega = \theta/\nu = 0.832(6)$ so $\theta = 0.524(4)$.

Leaving the pure FSS regime, now consider the overall temperature and
size dependence of $\chi(\beta,L)$.  Assuming $\beta_c$ known, the
critical exponent $\gamma_c$ can be estimated directly and
independently from an extrapolation to $\tau=0$ of the derivative
$\gamma_{\mathrm{eff}}(\tau,\infty) =
\partial\log\chi(\beta,L)/\partial\log\tau$ in the thermodynamic limit
conditions i.e. down to $L$-dependent crossover temperatures above
which the $\chi(\beta,L)$ are independent of $L$.  The crossover
occurs when $L \approx 6\xi(\beta,\infty)$, below which the
correlation length is no longer negligible compared to the sample
size. (As $T \to T_c$ below this crossover, $\chi(\beta,L)$ then tends
to a constant for each $L$).

There is obviously no "critical-to-classical crossover" as a function
of temperature. The crossover would appear automatically if the
effective exponent were defined (e.g. Ref.~\cite{luijten:97}) in terms
of the thermodynamic susceptibility
\begin{equation}
  \chi_{\mathrm{th}}(\beta) = [\partial m(\beta,h)/\partial h]_{h \to 0} \equiv
  \beta\chi(\beta)
  \label{chith}
\end{equation}
 and the traditional scaling variable $t$ through
\begin{equation}
  \gamma_{\mathrm{th,eff}} = \partial \log \chi_{\mathrm{th}}(\beta)/\partial \log t
  \label{gammath}
\end{equation}
because at high temperatures $\chi_{th}(\beta) \to \beta$ and $t \to T$.

The present data for $L=64$, $L=32$ and $L=16$ are of very high
statistical accuracy. Again assuming $\beta_c = 0.221655$,
$\gamma_{\mathrm{eff}}(\tau,L)$ values in the thermodynamic limit
conditions (which are in excellent agreement with HTSE data for
$\gamma_{\mathrm{eff}}(\tau,\infty)$ \cite{butera:02,butera:priv}),
can be extrapolated satisfactorily to $\tau=0$ assuming
$\gamma_{\mathrm{eff}}(\tau,\infty) = \gamma_{c} + a_{1}\tau^{\theta}
+\cdots $, Fig 3. The fit provides an estimate $\gamma = 1.239(1)$,
almost compatible with the HTSE \cite{butera:02} and FSS
\cite{deng:03} estimates.

\begin{figure}
  \includegraphics[width=3.5in]{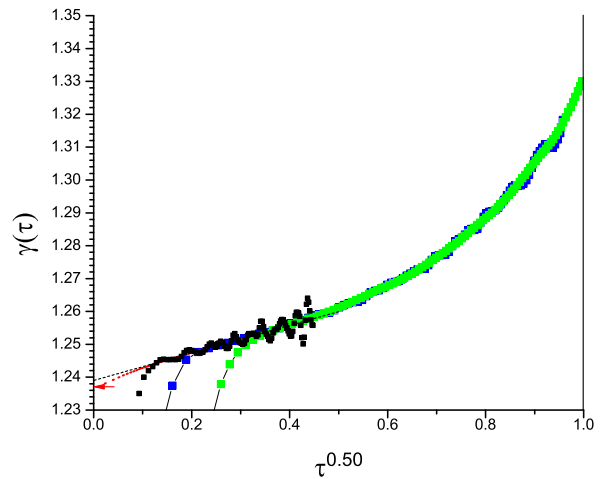}
  \caption{(Color online) An overall plot of the effective exponent
    $\gamma_{\mathrm{eff}}(\tau,L)$ fixing $\beta_c = 0.221655$ and
    $\theta = 0.50$ for sizes $L=64, 32, 16$ from top to bottom
    (black, blue, green). The thermodynamic limit envelope curve is
    clearly seen. The red line corresponds to an HTSE data analysis
    \cite{butera:02,butera:priv}, in full agreement with the present
    results over the entire temperature range except for a marginal
    difference near $\beta_c$. The red arrow indicates the consensus
    value for $\gamma(\beta_c)$.}  \protect\label{fig:3}
\end{figure}

\begin{figure}
  \includegraphics[width=3.5in]{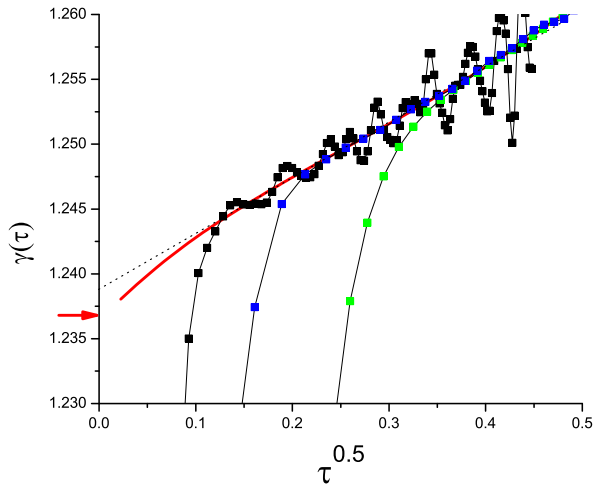}
  \caption{(Color online) As Fig 3, blowup of the small $\tau$
    region. } \protect\label{fig:4}
\end{figure}

The fluctuations in the plot for $L=64$ in Fig. 4 are an indication
of how sensitive these plots are to the slightest noise in the
original data. The temperature region in the far right of Fig. 4 for
$L=64$ corresponds to a region of energy levels measured at least
500,000 times. At the other end the energy levels were measured more
than 1,000,000 times. Data for still higher $L$ are not shown as the
fluctuations become more marked; unfortunately these higher $L$ data
cannot be used to refine the estimate of $\gamma$. The $\gamma$
estimate with the present method is sensitive to the value assumed for
$\theta$. The $\gamma$ estimate would become incompatible with the
consensus value if one assumed significantly higher values for
$\theta$, such as $0.54$ (estimates of $\theta$ are reviewed in
\cite{butera:02}).

An advantage of this $\gamma_{\mathrm{eff}}(\tau,\infty)$ technique is
that it is free from the problem of finite size corrections to
scaling, although the Wegner thermal corrections to scaling must be
taken into account as above. It can be noted also that this is a
direct measurement of $\gamma$ rather than an indirect estimate
through a combination of $\nu_c$ and $2-\eta_c$ estimates as is the
case for FSS.

\begin{figure}
  \includegraphics[width=3.5in]{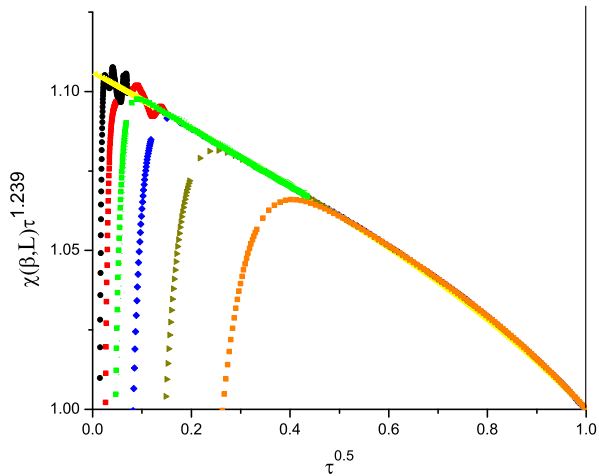}
  \caption{(Color online) The normalized susceptibility
    $\chi(\beta,L)\tau^{\gamma}$ against $\tau^{\theta}$ assuming
    $\gamma = 1.239$ and $\theta = 0.50$. Sizes $L=256, 128, 64, 32,
    16, 8$ from top to bottom (black, red, green, blue, olive,
    orange). The excellent fit (yellow) to the thermodynamic limit
    envelope data corresponds to Eq.(\ref{chiinfnorm}).  }
  \protect\label{fig:5}
\end{figure}

\begin{figure}
  \includegraphics[width=3.5in]{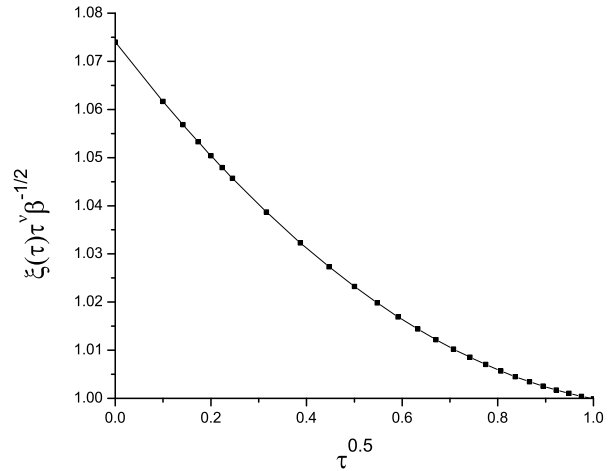}
  \caption{(Color online) The normalized correlation length
    $\xi(\beta,\infty)\tau^{\nu}\beta^{1/2}$ against $\tau^{\theta}$
    in the thermodynamic limit assuming $\nu = 0.630$ and $\theta =
    0.50$. Raw HTSE data provided by P. Butera
    \cite{butera:02,butera:priv}.}  \protect\label{fig:6}
\end{figure}

Fig. 5 shows the data for $L= 16$ to $L=256$ in the form of a
normalized plot, $\chi(\beta,L)\tau^{\gamma}$ against $\tau^{0.50}$
assuming $\gamma = 1.239$. Again it can be seen by inspection at which
point for each $L$ the curves leave the thermodynamic limit envelope
curve which is $L$ independent. With the scaling expression
Eq.(\ref{chiscal}) and using the data at the various $L$ but only in
the thermodynamic limit, the fit
\begin{equation}
  \chi(\beta,\infty)\tau^{\gamma} = 1.106\left(1 - 0.080\tau^{\theta}
  - 0.016\tau \right)
  \label{chiinfnorm}
\end{equation}
gives the values of the critical amplitude, $C_{\chi} = 1.106(5)$, and
the coefficient of the leading conformal and analytic correction
terms, $a_{\chi}=-0.080(3)$ and $b_{\chi}=-0.016(3)$, read directly
off the plot in Fig. 5. These values are fully consistent with but
more precise than earlier estimates from HTSE, $C_{\chi} = 1.11(1)$
and $a_{\chi}= -0.10(3)$ \cite{butera:98}, see \cite{campbell:08}. It
can be seen that the extended scaling expression with only two leading
Wegner correction terms gives a very accurate fit to the data over the
whole temperature range above the critical temperature.

If exactly the same data were expressed using $t=(T-T_c)/T_c$ as the
scaling variable rather than $\tau$, because $\tau = t/(1+t)$ one
would have to write
\begin{eqnarray}
  \chi(\beta,\infty) = ~~~~~~~~~~~~~~       \\   \nonumber
  1.106t^{-1.239}(1 + 1.239t + 0.1466t^2 - 0.0373t^3 + \cdots
  \\  \nonumber
  - 0.080t^{0.5}+ 0.0495t^{1.5}- 0.0371t^{2.5} +\cdots
  \\  \nonumber
  -0.016t + 0.016t^2 - 0.016t^3 +\cdots)
  \label{chiscalt}
\end{eqnarray}
Remembering that $t$ diverges at infinite $T$, each of the correction
terms in the sums is individually diverging at high temperatures.
Manifestly it is considerably more efficient to scale
$\chi(\beta,\infty)$ with $\tau$ rather than with $t$.

We have made no correlation length measurements. However we have
carried out an extended scaling parametrization of HTSE thermodynamic
limit second moment correlation length $\xi(\beta,\infty)$ data
supplied by P. Butera \cite{butera:02,butera:priv}.

\begin{figure}
  \includegraphics[width=3.5in]{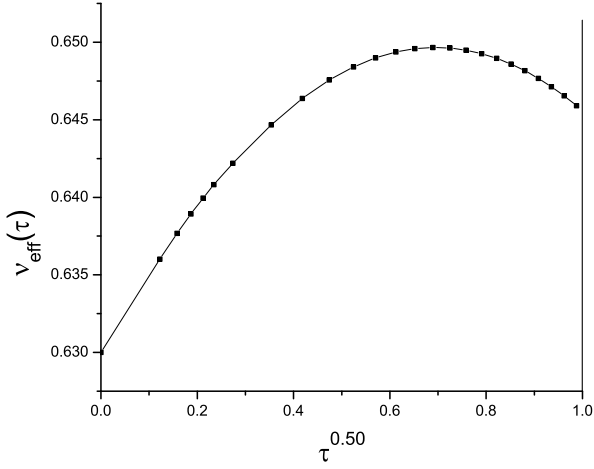}
  \caption{(Color online) The extended scaling effective exponent
    $\nu(\tau)$ against $\tau^{\theta}$ in the thermodynamic limit
    assuming and $\theta = 0.50$. Raw HTSE data provided by P. Butera
    \cite{butera:02,butera:priv} } \protect\label{fig:7}
\end{figure}

Fig. 6 shows a plot of the normalized correlation length
$\xi(\beta,\infty)\tau^{\nu}\beta^{1/2}$ against $\tau^{\theta}$
assuming $\nu = 0.630$ and $\theta = 0.50$. The data can be fitted
well by the extended scaling Wegner expression with two leading terms
only
\begin{equation}
  \xi(\beta,\infty)\tau^{\nu}\beta^{1/2} = 1.074\beta^{-1/2}\left(1 -
    0.120\tau^{0.5} + 0.051\tau \right)
  \label{xiinfnorm}
\end{equation}
(note that here the critical amplitude is $C_{\xi}/\beta_{c}^{1/2}$).
The same equation provides the temperature dependence of the effective
exponent defined by
\begin{equation}
  \nu_{\mathrm{eff}}(\beta,\infty) = \partial
  \log(\xi(\beta,\infty)/\beta^{1/2})/\partial \log\tau
\end{equation}
see Fig. 7. The effective exponent varies only by a few percent over
the whole range from $T=T_c$ to $T=\infty$. It is clear that the
$\beta^{1/2}$ prefactor is an essential part of the temperature
dependence of the correlation length.  The compact relation
Eq.(\ref{xiinfnorm}) is very useful as it allows finite size scaling
analyses of the entire data set for $\chi(\beta,L)$.

\section{Finite size scaling}

The extrapolation in Fig. 5 concerns only data in the thermodynamic
limit condition for each $L$. With Eq.(\ref{xiinfnorm}) in hand we can
plot all the data and not just the points in the thermodynamic limit
condition by appealing to the Privman-Fisher
relation~\cite{privman:84}, Eq.(\ref{PFFSS}).

As a first step we ignore corrections to scaling and draw, Fig. 8,
the leading order extended scaling FSS \cite{campbell:06} plot for the
susceptibility
\begin{equation}
  \chi(L,T)/(LT^{1/2})^{2-\eta} = F_{\chi}\left((LT^{1/2})^{1/\nu}\tau\right)
  \label{chiFSS}
\end{equation}
On the scale of the plot the scaling is already reasonable for all $T$
above $T_c$.

\begin{figure}
  \includegraphics[width=3.5in]{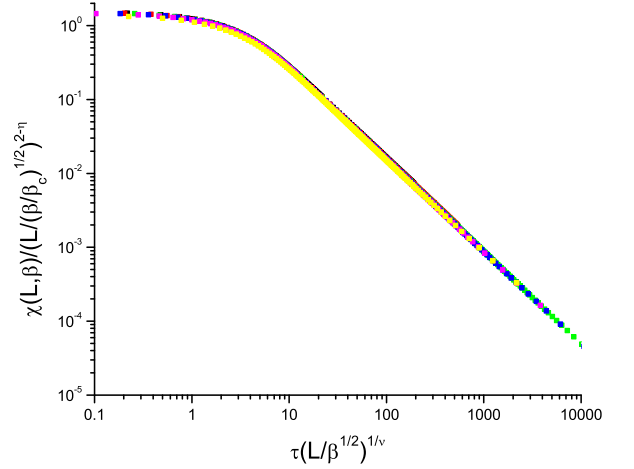}
  \caption{(Color online) The leading order extended scaling plot
    $\chi(L,T)/(L(T/T_c)^{1/2})^{2-\eta}$ against
    $\left((LT^{1/2})^{1/\nu}\tau\right)$ } \protect\label{fig:8}
\end{figure}

The conformal correction can then be introduced :
\begin{eqnarray}
  \chi(\beta,L)/\chi(\beta,\infty) \\   \nonumber
  = F_{\chi}\left(L/\xi(\beta,\infty)\right)(1 +
  a_{\chi}L^{-\omega}G_{\chi}\left(L/\xi(\beta,\infty)\right))
  \label{fishercorrchi}
\end{eqnarray}

The function $F(x)$ must have limits $F(x)\to 1$ at large $x$ and
$F(x) \sim x^{2-\eta}$ for small $x$. An explicit compact {\it ansatz}
which gives these limits automatically is
\begin{equation}
  F_{\chi}(x) =\left((1 - \exp(-bx^{(2-\eta)/a})\right)^a
  \label{Fchiform}
\end{equation}
where $x= L/\xi(\beta,\infty)$. In the critical limit $x \ll 1$,
\begin{equation}
F_{\chi}(x)=b^{a}(L/\xi(\beta,\infty))^{2-\eta}.
  \label{Fchicrit}
\end{equation}

By convention $G_{\chi}\left(0\right) = 1$.  Fig. 9 uses the
temperature dependence of the thermodynamic limit correlation length,
Eq.(\ref{xiinfnorm}), and the thermodynamic limit susceptibility,
Eq.(\ref{chiinfnorm}), to scale the data for all $L$ and all $\beta$
using Eq.(\ref{fishercorrchi}) for $\chi(\beta,L)$.

\begin{figure}
  \includegraphics[width=3.5in]{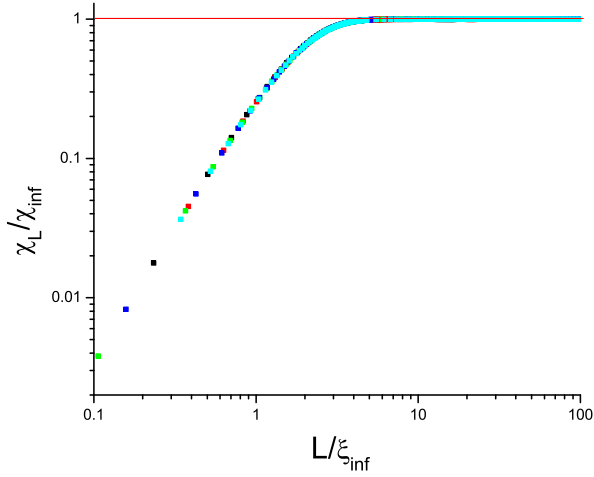}
  \caption{(Color online) The Privman-Fisher scaling plot
    $\chi(\beta,L)/\chi(\beta,\infty)$ against
    $L/\xi(\beta,\infty)$. $L= 256,128,64,32,16$, (black, red, green,
    blue,cyan) } \protect\label{fig:9}
\end{figure}

\begin{figure}
  \includegraphics[width=3.5in]{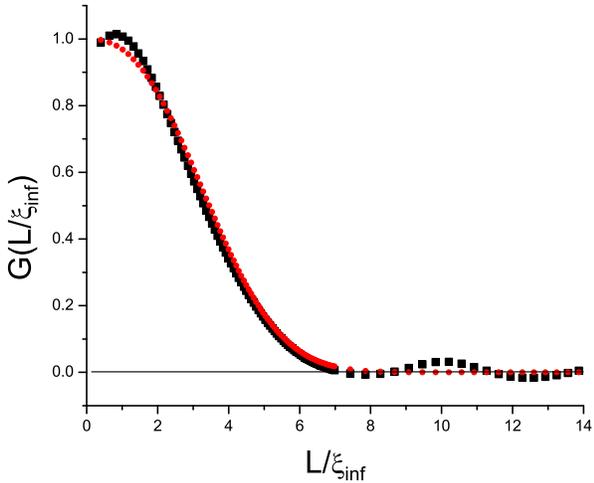}
  \caption{(Color online)The leading correction scaling function
    $G(L/\xi(\beta,\infty))$. Black squares: measured; red circles: fit } \protect\label{fig:10}
\end{figure}

The principle scaling function $F(x)$ and the leading
correction scaling function $G(x)$ were extracted from the
data. With the numerical constant $2-\eta$ fixed at $1.963$, an
accurate effective functional form for the principal scaling function
is
\begin{eqnarray}
F_{\chi}(x)=[1-\exp(-0.4179x^{1.963/1.262})]^{1.262}
  \label{FSSFchi}
\end{eqnarray}
On the scale of the figure $F(x)$ with these fit values ($a
= 1.262, b = 0.4179$) is indistinguishable from the overall curve
in Fig. 9.  By comparing data at small $L$ with data at large $L$
the correction to scaling function can also be estimated. A fit gives
$a_{\chi} \approx -0.22$ and
\begin{equation}
  G_{\chi}(x) \approx \exp\left(-0.038x^{2.5}\right)
  \label{FSSGchi}
\end{equation}
Fig. 10 shows the correction scaling function $G(x)$
together with the {\it ad hoc} Gaussian fit.

These FSS functions are universal to within metric constants \cite{fisher:72}.

In the same critical limit, from the definitions above
$\chi(\beta,\infty) = C_{\chi}\tau^{-\gamma}$ and $\xi(\beta,\infty)=
C_{\xi}\tau^{-\nu}$ so with $\chi(\beta_c,L) = C^{'}_{\chi}L^{2-\eta}$
in the large $L$ limit,
\begin{equation}
  b^a = C^{'}_{\chi}C_{\xi}^{2-\eta}/C_{\chi}
  \label{abCchi}
\end{equation}
The amplitudes $C^{'}_{\chi},C_{\chi}$ and $C_{\xi}$ are known from
critical and thermodynamic limit measurements respectively, so the
scaling form Eq.(\ref{Fchicrit}) has in principle only one free
parameter, $a$. Remarkably, when the other parameters are known, the
FSS crossover function can be encapsulated in one single parameter.

The overall scaling function expression covers all $L$ and all $T$
above $T_c$. The principle scaling function Eq.(\ref{Fchiform})
contains only one free parameter; it resembles the finite size scaling
form which has been used for the $2$d Villain model
\cite{katzgraber:08}. Previous expressions for principle finite
scaling functions \cite{caracciolo:95,kim:96}, in particular for the
$3$d Ising model \cite{kim:96}, were in the form of infinite series in
$\exp(-x)$ and so contained many fit parameters.

It would be of interest to study other members of the same family of
models in order to see if the compact form of scaling function
Eq.(\ref{Fchiform}) is generally valid, and how the universality is
expressed in the parameters $a$ and $b$.

Even below $T_c$ it has been noted that there should be a relationship
between the non-connected reduced susceptibility and the non-connected
correlation length \cite{hukushima:09}. The extended scaling gives
explicit leading order predictions for the asymptotic relations both
above and below $T_c$ between the finite size non-connected reduced
susceptibility $\chi(\beta,L)$ and the finite size non-connected
correlation length $\xi(\beta,L)$. As we have seen, in the limit
$\xi(\beta,L)/L \ll 1$
\begin{equation}
  \chi(\beta,L)/(LT^{1/2})^{2-\eta} \sim (\xi(\beta,L)/L)^{2-\eta}
\end{equation}
while in the opposite limit $\xi(\beta,L)/L \gg 1$ the predicted
relation is
\begin{equation}
  \chi(\beta,L)/(LT^{1/2})^{2-\eta} \sim (\xi(\beta,L)/L)^{(2/d)(d-2+\eta)}
\end{equation}
For the case of the $2$d square lattice Ising model the data confirm
both these relationships \cite{campbell:06,
  hukushima:09}. Unfortunately, as we have no data here for the finite
size $\xi(\beta,L)$ either above or below $T_c$ we cannot check the
relationship.

\section{Susceptibility above and below $T_c$}

The ratios of susceptibility amplitudes and of leading correction
factors above and below $T_c$ are universal. The standard reduced
susceptibility for the region above $T_c$ has been discussed; for $T$
above and below $T_c$ we will plot the modulus susceptibility
Eq.(\ref{chimoddef}) multiplied by $|\tau|^{\gamma}$ as a function of
$|\tau|^{\theta}$ with exponent values fixed at $\gamma = 1.239,
\theta = 0.50$, Fig. 11.

\begin{figure}
  \includegraphics[width=3.5in]{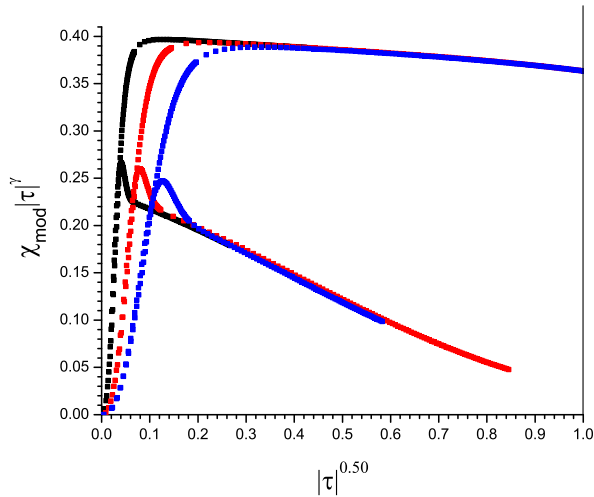}
  \caption{(Color online) The normalized modulus susceptibility
    $\chi_{\mathrm{mod}}(\tau)|\tau|^{\gamma}$ as a function of
    $|\tau|^{\theta}$. The upper set of curves corresponds to $T >
    T_c$ and the lower set to $T < T_c$. In both cases the sizes
    are $L=64,32,16$ (black, red, blue).}  \protect\label{fig:11}
\end{figure}
By definition $\chi_{\mathrm{mod}}$ becomes equal to the connected
reduced susceptibility below $T_c$ in the thermodynamic large $L$
limit. Extrapolating the data corresponding to this limit to $|\tau| =
0$ we find to leading order
\begin{equation}
  \chi_{\mathrm{conn}} = C_{\chi,-}|\tau|^{-1.239}\left(1 +
    a_{\chi,-}|\tau|^{0.50} + \cdots\right)
  \label{chiconn}
\end{equation}
with $C_{\chi,-} = 0.241(2)$ and $a_{\chi,-}= -0.82(5)$. Taking
into account the normalization factor for $\chi_{\mathrm{mod}}$, the
present estimates for the amplitude ratio and the correction amplitude
ratio are $C_{\chi,+}/C_{\chi,-} = 4.67(3)$ and
$a_{\chi,+}/a_{\chi,-} = 0.111(10)$. The amplitude ratio is
consistent with previous Monte-Carlo estimates, $4.75(3),4.72(11)$ and
$4.713(7)$, \cite{caselle:97,engels:99,hasenbusch:10}. The present
correction amplitude ratio estimate is however significantly lower
than a field theory value $0.315(13)$ \cite{bagnuls:87,privman:91}.

\section{Specific heat above and below $T_c$}

The specific heat is intrinsically difficult to analyze because of the
strong regular term $C_0$ and the small value of the critical exponent
$\alpha$ (see Eq.(\ref{Cvextdef})). It turns out in addition that
there are strong and peculiar finite size corrections. On the other
hand the statistical precision of the specific heat data is very high;
data for $L=512$ were included in this analysis.  The general leading
form of the envelope data in the thermodynamic limit condition is
assumed to be $\left(C_v(\beta,L) - C_0\right)|\tau_{2}|^{\alpha} =
C_{c}\left(1 + a_{c}|\tau_{2}|^{\theta}\right)$ where $\tau_{2} =
1-(\beta/\beta_c)^2$.  The amplitudes are $C_{c,+}, C_{c,-}$ and
$a_{c,+}, a_{c,-}$ above and below $T_c$ respectively. Here $\alpha$
is fixed at $0.110$, which is the expected value from the relation
$\alpha = \nu d - 2$ with $\nu = 0.630$. The regular term $C_0$ is
assumed to be temperature independent; the estimate $C_0= -30.9$ is
obtained from the overall fit discussed below. It should be underlined
that the extended scaling variable is $\tau_2$ not $\tau$.

In the high temperature range (down to $\beta \approx 0.2$) the data
can be compared to data points derived by directly summing the HTSE
terms up to $n=46$ from Ref.~\cite{arisue:03}. Point by point
agreement is better than to $1$ part in $10^3$.

As a first step we plot the raw $\log C_{v}(\tau_2,L)$ above $T_c$
against $\log\tau_{2}$, Fig. 12.  The thermodynamic limit data for
different $L$ can be clearly observed but the points fall on a curve
rather on a straight line even down to very small $\tau_{2}$; this is
because no $C_{0}$ term has been allowed for. Next, we plot
$\log\left(C_v(\tau_{2},L)-C_{0}\right)$ against $\log\tau_{2}$, as
in Fig. 13 for various trial values of $C_{0}$. In Fig. 13 with
$C_0= -30.9$ the envelope data now lie on a straight line of slope
$-0.110$ for the lower range of $\tau_{2}$ (and the larger $L$). We
make a Privman-Fisher finite size scaling plot of
$\left(C_v(\tau_{2},L)+30.9\right)/\left(C_v(\tau_{2},\infty)+30.9\right)$
against $L/\xi(\tau_{2},\infty)$ with
$\left(C_v(\tau_{2},\infty)+30.9\right)$ taken from the extrapolated
envelope for small $\tau_{2}$ and the measured envelope curve for
higher $\tau_{2}$, fitted to an explicit function for
$\left(C_v(\tau_{2},\infty)+30.9\right)$, Fig. 14. The thermodynamic
limit correlation length is taken from Eq.(\ref{xiinfnorm}).

In the finite size limited $L < \xi(\tau_{2},\infty)$ region the
normalized specific heat shows a strong peak, in contrast to the
regular FSS crossover observed for the susceptibility.  The quality of
the global fit is sensitive to the value chosen for the regular term
$C_0$ as the correct choice for this parameter is essential to obtain
an $L$-independent peak height in Fig. 13.  Once $C_0$ is fixed fine
adjustments are made to the correction terms so as to obtain an $L$-
and $T$-independent flat plateau in the left hand side thermodynamic
limit region.

An excellent global  Eq.(\ref{PFFSS}) FSS fit is obtained taking
\begin{eqnarray}
  C_v(\tau_{2},\infty) =  ~~~~~~~~~~~~~~~~~~~~~~~~~\\  \nonumber
  -30.9 + 29.85\tau_{2}^{-0.11}\left(1 + 0.12\tau_{2}^{0.5} + 0.014\tau_{2}\right)
  \label{3dCv}
\end{eqnarray}

The optimal value $C_0 = -30.9(5)$ can be compared with previous
estimates : $-33.3(24)$ \cite{hasenbusch:98} and $-27.85(80)$
\cite{feng:10}.

The normalized $(C_v(\tau_{2},\infty)+30.9)\tau_{2}^{0.11}$ is shown
in Fig. 15 where the nearly linear thermodynamic limit envelope is
obvious.

\begin{figure}
  \includegraphics[width=3.5in]{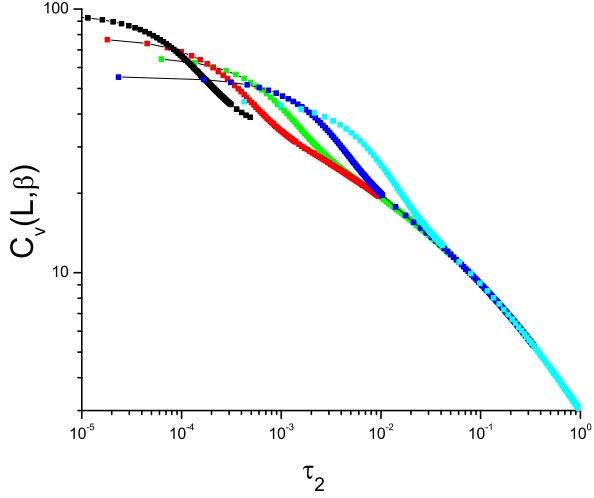}
  \caption{(Color online) Raw $\log C_v(\beta,L)$ against
    $\log\tau_{2} = \log(1-(\beta/\beta_c)^2)$. The sizes are from
    left to right $L = 512, 256, 128, 64, 32$, (black, red, green,
    blue, cyan) } \protect\label{fig:12}
\end{figure}

\begin{figure}
  \includegraphics[width=3.5in]{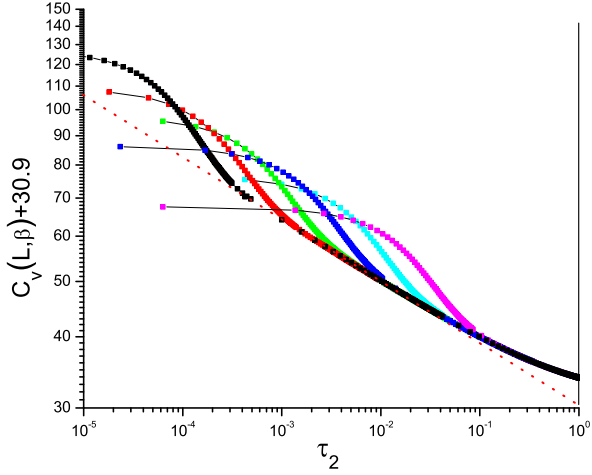}
  \caption{(Color online) $\log(C_v(\beta,L)+ 30.9)$ against $\log
    \tau_{2}$. The sizes are from left to right $L = 512, 256, 128,
    64, 32, 16$, (black, red, green, blue, cyan, magenta). The dashed
    line has the slope $-0.110$.} \protect\label{fig:13}
\end{figure}

\begin{figure}
  \includegraphics[width=3.5in]{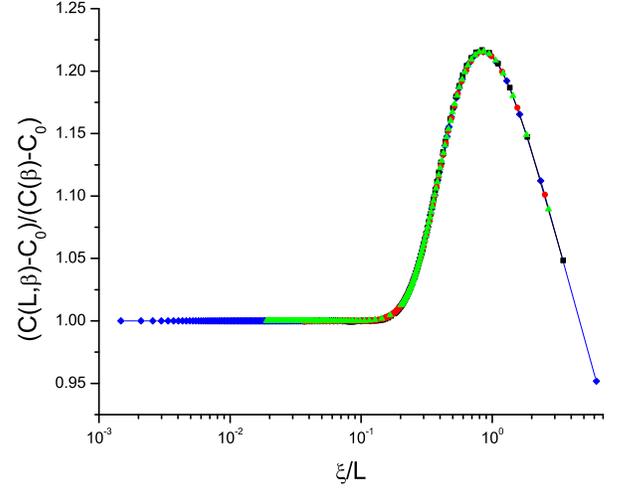}
  \caption{(Color online) The specific heat finite size scaling
    fit. The ratio $\left(C_v(\beta,L) -
    C_0\right)/\left(C_v(\beta,\infty) - C_0\right)$ against
    $\xi(\beta,\infty)/L$ with $C_0 = -30.9$. Sizes are $L = 512, 256,
    128, 64$, (black, red, green, blue)} \protect\label{fig:14}
\end{figure}

\begin{figure}
  \includegraphics[width=3.5in]{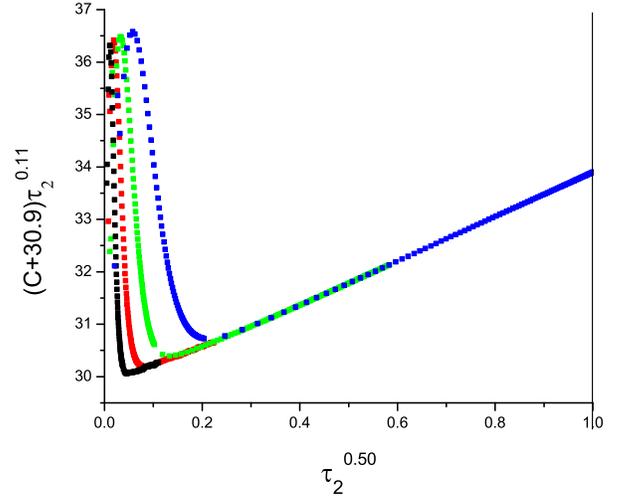}
  \caption{(Color online) $\left(C_v(\beta,L) -
    C_0\right)\tau_{2}^{\alpha}$ against $\tau_{2} =
    1-(\beta/\beta_c)^2$ with $\alpha = 0.110$ and $C_0 = -30.9$ for
    all temperatures $T > T_c$. Sizes $L=256, 128, 64, 32$, (black,
    red, green, blue).}  \protect\label{fig:15}
\end{figure}

\begin{figure}
  \includegraphics[width=3.5in]{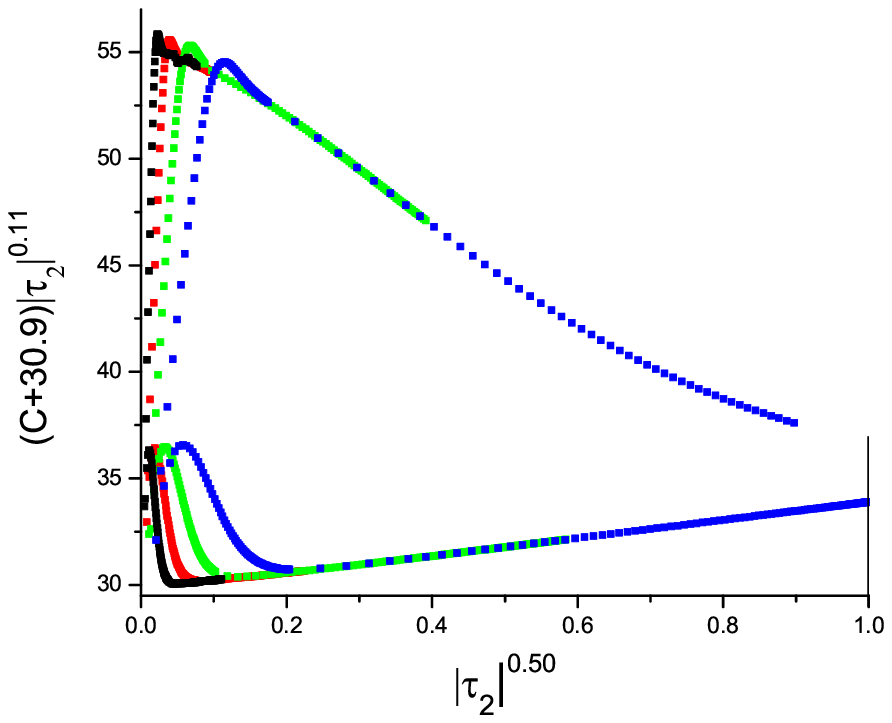}
  \caption{(Color online) $\left(C_v(\beta,L) -
    C_0\right)|\tau_{2}|^{\alpha}$ against $|\tau_{2}| =
    |1-(\beta/\beta_c)^2|$ with $\alpha = 0.110$ and $C_0 =
    -30.9$. The lower set of curves corresponds to $T > T_c$ and the
    upper set to $T < T_c$. Sizes $L=256, 128, 64, 32$, (black, red,
    green, blue).}  \protect\label{fig:16}
\end{figure}

The $C_{v}(\beta,\infty)$ from Eq.(\ref{3dCv}) together with the
peaked FSS curve (for which we have no explicit algebraic expression)
provide an accurate representation of the specific heat at all
temperatures above $T_c$ and for all sizes $L$. This is in contrast to
previous analyses of MC data which were made in terms of truncated
series of terms.

The ratios of critical amplitudes and of leading correction amplitudes
above and below $T_c$ are universal. The data show critical amplitudes
$C_{c,+} = 29.9(1)$ and $C_{c,-} = 55.4(2)$ above and below $T_c$,
Fig. 16.  With the extended scaling definition, $C_{c,+} =
A_{c}2^{\alpha}/\beta_{c}^{2}$ where $A_{c}$ is the amplitude using
the standard definition. The present $C_{c,+}$ result is in very good
agreement with the HTSE estimate $A_{c}=1.34(1)$ given in
Ref.~\cite{butera:02} which corresponds to $C_{c,+} = 29.4(3)$. The
present estimate for the amplitude ratio (which is definition
independent) is $C_{c,+}/C_{c,-} = 0.540(4)$, consistent with
$\epsilon$-expansion and field theory values of $0.524(10)$ and
$0.541(14)$ respectively \cite{privman:91}, and with the most recent
MC values $0.532(7)$ and $0.536(2)$ \cite{feng:10,hasenbusch:10}. For the correction amplitudes the
data indicate (Fig. 15 and 16) $a_{c,+} \approx 0.12$ and $a_{c,-}
\approx -0.23$, so $a_{c,+}/a_{c,-} \approx -0.52$ and
$a_{c,+}/a_{\chi,+} \approx 1.4$. These values can be compared with
field theory estimates, $0.96(25)$ and $0.95(10)$ respectively
\cite{bagnuls:81,bagnuls:87,privman:91}. (It should be noted that the
$a_{c}$ values in our notation correspond to $a_{c}\alpha$ in the
notation of Refs.~ \cite{bagnuls:81,bagnuls:87}.)

We cannot carry out a full FSS analysis below $T_c$ as we lack
information on the correlation length.

\section{Conclusion}

We have applied the extended scaling approach to the analysis of two
canonical Ising ferromagnet models : the historic $S=1/2$ ferromagnet
on a $1$d chain, and the $S=1/2$ ferromagnet on the simple cubic
lattice.

For the $1$d model, with the scaling variables $\tau=(1-\tanh\beta)$
for the susceptibility and the correlation length and $\tau_{2} =
(1-\tanh\beta)^2$, all the analytic thermodynamic limit expressions
are of precisely the extended scaling form over the entire temperature
range from zero to infinity, with no confluent corrections,
Eqns.~\ref{chiext1d}, \ref{xiext1d}, \ref{Cvext1d}.

An appropriate scaling variable for reduced susceptibility and second
moment correlation length in a ferromagnetic Ising model with a
non-zero ordering temperature is $\tau = 1 - \beta/\beta_c =
(T-T_c)/T$, not the traditional $t = (T-T_c)/T_c$.  An exhaustive
analysis of high quality numerical data for the $3$d Ising model
demonstrates that the reduced susceptibility and the second moment
correlation length can be represented satisfactorily over the entire
temperature range above $T_c$ by compact expressions containing two
leading Wegner correction terms only :
\begin{equation}
  \chi(\beta,\infty)= 1.106\tau^{-1.239}\left(1 - 0.080\tau^{0.5} -
  0.016\tau \right)
  \label{chinormbis}
\end{equation}
and
\begin{equation}
  \xi(\beta,\infty) = 1.074\beta^{1/2}\tau^{-0.630}\left(1 -
    0.109\tau^{0.5} + 0.039\tau\right)
  \label{3dxibis}
\end{equation}

For the specific heat on a bipartite lattice (such as the sc lattice)
the appropriate extended scaling variable is $\tau_{2} = 1 -
(\beta/\beta_c)^2$. The data from $T_c$ to infinite temperature can be
fitted accurately by
\begin{eqnarray}
  C_v(\beta,\infty) = ~~~~~~~~~~~~~~~~~~~~~~~~~~~~~~~ \\   \nonumber
  -30.9 + 29.85\tau_{2}^{-0.110}\left(1 + 0.12\tau_{2}^{0.5}+ 0.014\tau_{2}\right)
  \label{3dCv2}
\end{eqnarray}

We give explicit finite size susceptibility scaling functions for the
two models. The principle $1$d susceptibility scaling function
\begin{equation}
\chi(\beta,L)/\chi(\beta,\infty)= \tanh(L/2\xi(\beta,\infty))
\end{equation}
is exact. The principle $3$d susceptibility scaling {\it ansatz}
\begin{equation}
\chi(\beta,L)/\chi(\beta,\infty)=[1 -\exp(-b(L/\xi(\beta,\infty))^{(2-\eta)/a})]^a
\end{equation}
with $a=1.262$, $b=0.4179$ fits the data to high precision. This
form where two parameters encapsulate the finite size scaling
crossover from the region $L \gg \xi(\beta,\infty)$ to the region $L
\ll \xi(\beta,\infty)$ might well be of generic application.

The critical parameters can be estimated by combining the data in the
thermodynamic limit $L \gg \xi_{\infty}(\beta)$ with the data in the
finite size scaling region $L \ll \xi_{\infty}(\beta)$.  The results
provide complementary estimates for critical amplitudes and critical amplitude ratios.

The aim of this work is however not so much to improve on the already
very accurate existing estimates for universal critical parameters in
the intensively studied ferromagnetic $3$d Ising model, but to explain
the rationale leading to an optimized choice of scaling variables and
scaling expressions for covering the whole temperature range up to
infinite temperature. Here we spell out in detail for two canonical
examples, the $1$d and $3$d Ising ferromagnets, an "extended scaling"
methodology for studying numerical data taken over the entire
temperature range without restricting the analysis to a narrow
"critical" temperature region near $T_c$. Scaling variables and
scaling expressions are chosen following a simple unambiguous
prescription inspired by the well established HTSE approach. Using
these and allowing for small leading Wegner correction terms where
necessary, critical scaling expressions for $\chi(\beta,\infty)$,
$\xi(\beta,\infty)$ and $C_{v}(\beta,\infty)$ remain valid to high
precision from $T_c$ right up to infinite temperature. Residual
analytic correction terms are either strictly zero (in $1$d) or very
weak (in $3$d).

The approach can readily be generalized to other less well understood systems.

\section {Appendix A : General spin S}

Standard expressions for the reduced susceptibility and the
correlation length for ferromagnets as defined in \cite{butera:02} are
for general spin $S$
\begin{equation}
  \chi(\tau) = C_{\chi}^{S}(std)\tau^{-\gamma}\left(1 + F_{\chi}(\tau)\right)
  \label{Cchibutera}
\end{equation}
and
\begin{equation}
  \xi(\tau) = C_{\xi}^{S}(std)\tau^{-\nu}\left(1 + F_{\xi}(\tau)\right)
  \label{Cxibutera}
\end{equation}

The extended scaling prescription consists in transposing each HTSE
expression such that it takes the form of a series in a variable $x$,
having leading term $1$ and multiplied by a prefactor. In the case of
a finite critical temperature Ising ferromagnet with
$\tau=1-\beta/\beta_c$, the critical amplitudes are then defined
through
\begin{equation}
  \chi(\beta,\infty) = C_{\chi}(es)\tau^{-\gamma}\left( 1 + F_{\chi}(\tau) \right)
  \label{chiesdef}
\end{equation}
(c.f. Eq.(\ref{Qq})) and
\begin{equation}
  \xi(\beta,\infty) = C_{\xi}(es)\beta^{1/2}\tau^{-\nu}\left( 1 +
  F_{\xi}(\tau)\right)
  \label{xiesdef}
\end{equation}

For general Ising
spin $S$, dimension $d$, and a lattice with $z$ nearest neighbors the
extended scaling critical amplitudes are
\begin{equation}
  C_{\chi}^{S}(es) = \left((S+1)/3S\right)C_{\chi}^{S}(std)
  \label{Cchiext}
\end{equation}
and
\begin{equation}
  C_{\xi}^{S}(es) = C_{\xi}^{S}(std)/\left(z(1+S)\beta_c/6dS\right)^{1/2}
  \label{Cxiext}
\end{equation}

The definitions of the effective exponents are unaltered.

With these normalizations the physical significance of the critical
amplitudes becomes much more transparent. Ref.~\cite{butera:02} lists
the standard critical amplitudes as functions of $S$ for sc and bcc
lattices. In Table AI we compare these values with those obtained
using the above definitions. The extended scaling values are close to
$1$ for all $S$; the differences $\left(C_{Q}^{S}(es)-1\right)$ which
can be read directly from the Table are a quantitative indication,
model by model, of the amplitude of the $S$-dependent correction terms
within $F_{Q}^{S}$.

\begin{table}[htbp]
  \caption{\label{Table:AI} Values of the critical amplitudes for spin
    $S$ with the standard definitions, Eqns.~\ref{Cchibutera} and
    \ref{Cxibutera} Ref.~\cite{butera:02} compared with values using
    the extended scaling definitions Eqns.~\ref{Cchiext} and
    \ref{Cxiext}}
  \begin{ruledtabular}
\begin{tabular}{cccccccc}

$S$&$1/2$&$1$&$3/2$&$2$&$5/2$&$3$&$\infty$ \\
$sc$&&&&&&&   \\
$C_{\chi}(std)$&$1.127$&$0.682$&$0.545$&$0.482$&$0.443$&$0.418$&$0.307$ \\
$C_{\xi}(std)$&$0.506$&$0.458$&$0.443$&$0.436$&$0.432$&$0.430$&$0.423$ \\
&&&&&&&  \\
$C_{\chi}(es)$&$1.127$&$1.023$&$0.981$&$0.964$&$0.949$&$0.941$&$0.922$ \\
$C_{\xi}(es)$&$1.075$&$1.003$&$0.979$&$0.967$&$0.960$&$0.957$&$0.945$ \\
&&&&&&&  \\
$bcc$&&&&&&&   \\
$C_{\chi}(std)$&$1.042$&$0.622$&$0.497$&$0.438$&$0.404$&$0.383$&$0.282$ \\
$C_{\xi}(std)$&$0.469$&$0.426$&$0.411$&$0.405$&$0.401$&$0.399$&$0.394$ \\
&&&&&&&  \\
$C_{\chi}(es)$&$1.042$&$0.933$&$0.894$&$0.876$&$0.867$&$0.861$&$0.845$ \\
$C_{\xi}(es)$&$1.023$&$0.953$&$0.927$&$0.915$&$0.909$&$0.905$&$0.895$ \\

\end{tabular}
\end{ruledtabular}
\end{table}

If the corrections to scaling up to infinite temperature are dominated
by the leading (confluent) term then $C_{\chi}^{S}(es)-1
\approx a_{\chi}(S)$ and $C_{\xi}^{S}(es)-1 \approx a_{\xi}(S)$. The
universal ratio $a_{\xi}(S)/a_{\chi}(S) \approx
(C_{\xi}^{S}(es)-1)/(C_{\chi}^{S}(es)-1)$. From Fig. 13 which shows the data
from the Table, we can estimate $a_{\xi}(S)/a_{\chi}(S) \approx
0.65(2)$ (with a small offset corresponding to the next-to-leading
correction). This compares favorably with the estimates $0.76(6)$ from
HTSE \cite{butera:02}, $0.65(5)$ obtained by the RG in the
perturbative fixed-dimension approach at sixth order
\cite{bagnuls:86}, and $0.65$ from the $\epsilon$ expansion to second
order \cite{chang:83}.

\begin{figure}
  \includegraphics[width=3.5in]{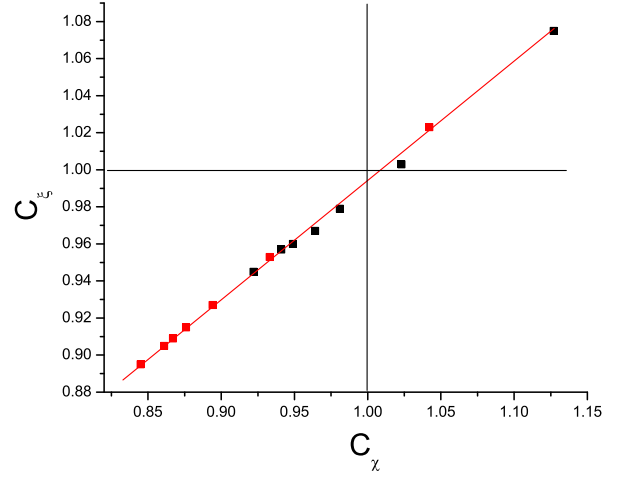}
  \caption{(Color online) $C_{\xi}^{S}(es)$ plotted against $C_{\chi}^{S}(es)$,
    where $C_{\xi}^{S}(es)$ and $C_{\chi}^{S}(es)$ are spin $S$ dependent extended
    scaling susceptibility and correlation length critical
    amplitudes. Black points sc lattice, red points bcc lattice. See
    text, Ref.~\cite{butera:02}, and Table AI. }
  \protect\label{fig:17}
\end{figure}

\section {Appendix B: Ising spin glass}

It can be noted that in the case of the Ising Spin Glass the energy
scale of the interactions is fixed by $\langle J_{ij}^2 \rangle$ not
by $\langle J \rangle$ as in the ferromagnetic case ($\langle J_{ij}
\rangle$ is zero in a symmetric interaction distribution spin
glass). From an obvious dimensional argument the normalized spin glass
"temperature" should be $T^2/\langle J_{ij}^2 \rangle$. It has long
been recognized that for the spin glass the HTSE expressions contain
even terms only (i.e. an expansion in $(\beta/\beta_c)^{2}$ or
$(\tanh\beta/\tanh\beta_c)^2$ rather than in $\beta/\beta_c$) so
the appropriate scaling variable is $\tau_{sg} = 1 -(\beta/\beta_c)^2$
or $1-(\tanh\beta/\tanh\beta_c)^2$
\cite{fisch:77,singh:86,klein:91,daboul:04}. The argument presented
above for the ferromagnet can be repeated {\it mutatis mutandis} on
this basis; the extended scaling expressions for $\chi(\beta,L)$ and
$\xi(\beta,L)$ in spin glasses are the same as those for the
ferromagnet (Eqns.~\ref{chiextdef} and ~\ref{xiextdef}) but with
$(\beta/\beta_c)^2$ substituted for $\beta/\beta_c$ everywhere
\cite{campbell:06}.

Unfortunately the great majority of publications on spin glasses have
used $t$ as the scaling variable which is quite inappropriate except
for a very restricted range of temperatures near $T_c$.  One
consequence is that many published estimates of the exponent $\nu$ in
spin glasses are low by a factor of about $2$ (see the discussion in
\cite{katzgraber:06}).

\section{Acknowledgements}
We would like to thank Paolo Butera for generously providing us
with tabulated data sets and for helpful comments. We thank
V. Privman for an encouraging comment. This research was conducted
using the resources of High Performance Computing Center North
(HPC2N).

{99}

\end{document}